\documentclass[sigconf, authorversion,nonacm]{acmart}
\AtBeginDocument{%
  \providecommand\BibTeX{{%
    \normalfont B\kern-0.5em{\scshape i\kern-0.25em b}\kern-0.8em\TeX}}}

\usepackage{soul, xcolor}

\begin{document}

\title{Asynchronous Neuromorphic Optimization with Lava}

\author{Shay Snyder}
\email{ssnyde9@gmu.edu}
\orcid{0000-0002-3369-3478}
\affiliation{%
  \institution{George Mason University}
  \streetaddress{4400 University Dr}
  \city{Fairfax}
  \state{Virginia}
  \country{USA}
  \postcode{22030}
}
\author{Sumedh R. Risbud}
\email{sumedh.risbud@intel.com}
\orcid{0000-0003-4777-1139}
\affiliation{%
  \institution{Intel Labs}
  \streetaddress{2200 Mission College Blvd}
  \city{Santa Clara}
  \state{California}
  \country{USA}
  \postcode{95054}
}
\author{Maryam Parsa}
\email{mparsa@gmu.edu}
\orcid{0000-0002-4855-4593}
\affiliation{%
  \institution{George Mason University}
  \streetaddress{4400 University Dr}
  \city{Fairfax}
  \state{Virginia}
  \country{USA}
  \postcode{22030}
}

\renewcommand{\shortauthors}{Snyder, et al.}


% \received{12 April 2024}

\begin{abstract}
    Performing optimization with event-based asynchronous neuromorphic systems presents significant challenges. Intel's neuromorphic computing framework, Lava, offers an abstract application programming interface designed for constructing event-based computational graphs. In this study, we introduce a novel framework tailored for asynchronous Bayesian optimization that is also compatible with Loihi 2.
    % Our framework seamlessly integrates with Loihi 2 and supports three search algorithms: Bayesian Optimization, grid search, and random search.
    % \textcolor{blue}{We showcase the capability of our system with a test scenario where Bayesian Optimization is running on the CPU and asynchronously communicating with a spiking convolutional neural network running on physical Loihi hardware.}
    We showcase the capability of our asynchronous optimization framework by connecting it with a graph-based satellite scheduling problem running on physical Loihi 2 hardware.
\end{abstract}

\maketitle

\section{Introduction}
True neuromorphic systems operate using an event-driven architecture that is radically different from traditional von-Neumann based systems.
Transitioning to event-based communication requires a drastic shift in programmatic design and theoretic analysis. Similar to modern networking systems that need to operate when packets get lost or corrupted~\cite{djahel2010mitigating}, resilient neuromorphic systems should properly handle when spikes, or other forms of information, do not arrive at the pre-specified interval.

Intel's neuromorphic computing framework, Lava~\cite{lava}, bridges the gap between existing sequential computational graphs and translates them into heterogeneous, asynchronous structures. A variety of algorithms have been implemented within Lava such as Gaussian process regression in Lava Bayesian Optimization (Lava BO)~\cite{10.1145/3589737.3605998} and Quadratic Unconstrained Binary Optimization~\cite{lava}. However, these solvers and optimization algorithms do not have the infrastructure to support event-based communication when problems are being executed on separate compute nodes or architectures. This causes deadlocking issues and wasted CPU clock cycles as processes wait until information is received.

In this study, we introduce a novel framework tailored for asynchronous Bayesian optimization within Lava.
% Our framework seamlessly integrates with Loihi 2 and supports three search algorithms: Bayesian Optimization~\cite{10.1145/3589737.3605998}, grid search~\cite{bergstra2012random}, and random search~\cite{bergstra2012random}.
% While we have included a series of predefined implementations, our approach is scalable to any other optimization or search algorithm that will be implemented within future versions of the Lava software framework.
Our framework seamlessly integrates with Loihi 2 and is scalable to any other optimization or search algorithm that will be implemented within future versions of the Lava software framework.
%\textcolor{blue}{We showcase the capability of our system with a deep spiking convolutional neural network for NMNIST classification running on Loihi 2~\cite{lava} hardware.}
We showcase ability of our system to asynchronously communicate with a quadratic unconstrained binary solver applied to a graph-based satellite scheduling problem running on Loihi 2.

\section{System Architecture}

Individual computational elements in Lava~\cite{lava} are represented as (\textit{Processes}) that provide an abstract blueprint for input ports, output ports, internal variables, and variable reference ports. Taking a specific \textit{Process} and implementing the underlying behavior for specific computing architectures are \textit{Process Models}. Therefore, \textit{Process Models} allow a single type of process, such as a Leaky-Integrate-and-Fire neuron, to be executed on different compute platforms\footnote{See \url{http://lava-nc.org} for details about Lava concepts like \textit{Process} and \textit{Process Models}}. For example, it could be running on a central processing unit with floating point precision or a Loihi 2 neurocore with integer precision.
% such as central processing units with floating-point precision or fixed, integer-precision architectures such as Loihi 2 neurocores

\begin{figure*}
    \centering
    \includegraphics[width=\textwidth]{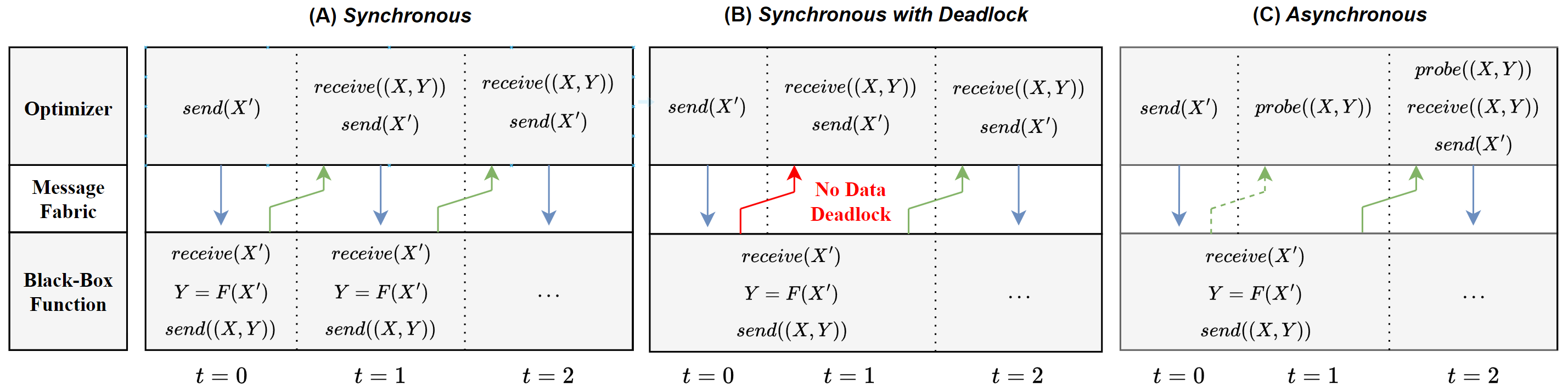}
    \caption{State diagrams comparing optimizer to black-box function communication with synchronous and asynchronous operation. (A) Synchronous communication where the optimizer transmits a single hyperparameter value $X'$. The black-box function receives $X'$ and calculates $Y$ via $Y=F(X')$ before returning the tuple $(X,Y)$. (B) The same scenario as A but where the black-box function takes multiple steps to calculate the result $Y$. This causes a programmatic deadlock where the optimizer is receiving data that isn't available. (C) Using a probe, the optimizer will determine if the result is fully processed from the black-box function. If the data isn't available, the process will proceed to the next timestep.}
    \label{fig:state-diagram}
\end{figure*}

Different processes are coupled by connecting the output port of one process to the input port of another process. For example, the output port of a cluster of LIF neurons is connected to the input port of a dense layer. In more complicated process graphs, such as iterative optimization and search algorithms, port connections can span between different hardware elements such as a neurocore sending output spikes to the embedded CPU or super host. This cross device communication presents numerous challenges where the determinism of synchronous architectures is lost. Having multiple processes communicate while running on different architectures leads to issues with deadlocking and excess computation. Deadlocking occurs when a port is trying to receive information but there is no data available. This will cause the process to freeze indefinitely and waste processor clock cycles by not allowing other threads to execute. A visual demonstration of synchronous and asynchronous communication between the optimizer and black-box function is shown in Figure~\ref{fig:state-diagram}. Herein, the cause of deadlocking within synchronous communication is highlighted along with the corresponding alleviation of this issue with the transition to an asynchronous paradigm.

\begin{figure}[ht]
    \centering
    \includegraphics[width=0.44\textwidth]{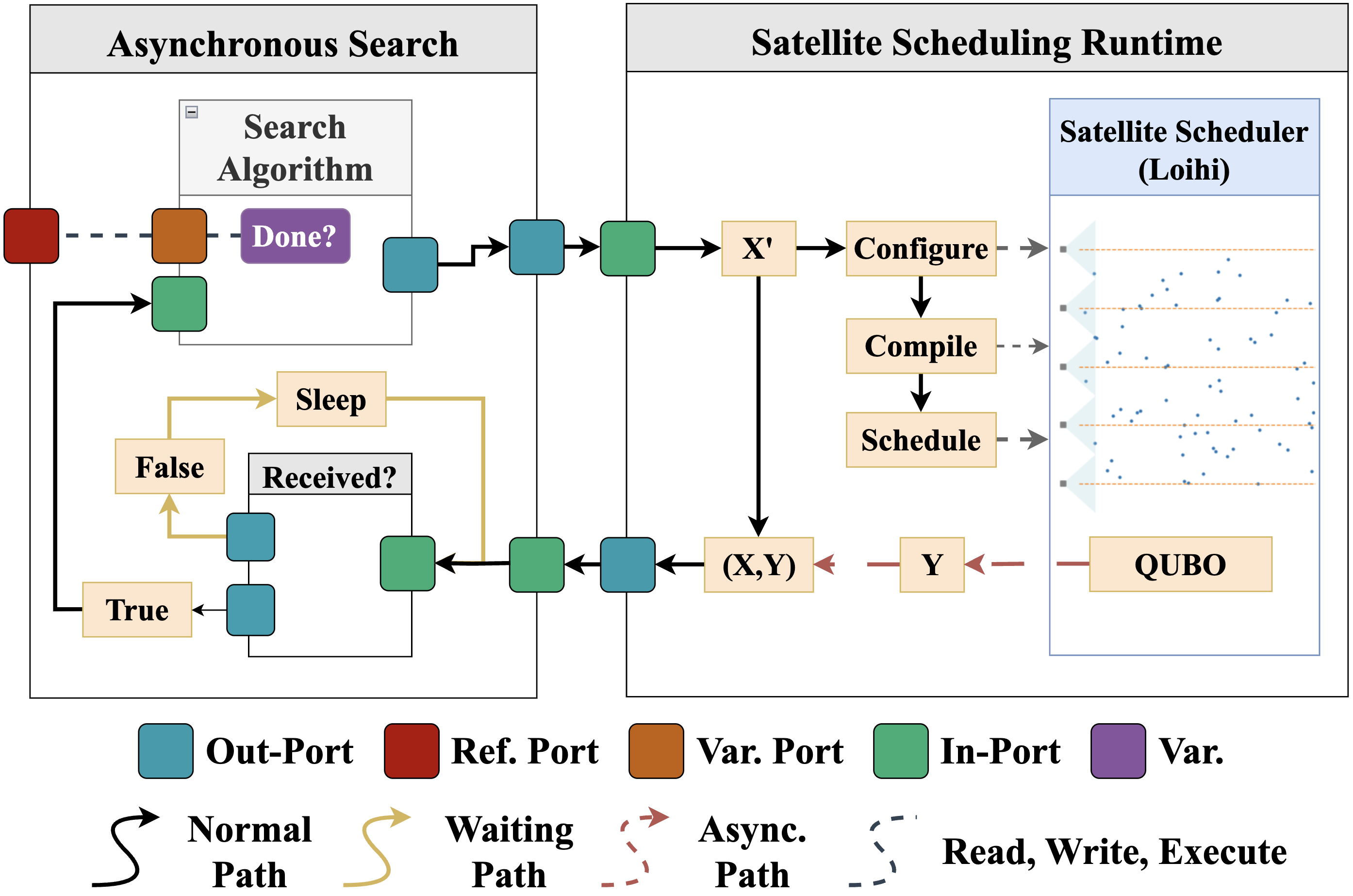}
    % \caption{The communication loop between our asynchronous optimization framework in Lava and an abstract black-box function. \textbf{$X'$} is an unknown set of parameters from the search algorithm where \textbf{$f(x')$} is function of the parameters. In the case of Loihi, this function would transfer the parameters to the neurocore, run the neurocore for a prespecified number of steps, and transfer the outputs back to the CPU. \textbf{(X, Y)} the parameter configuration along with the corresponding output. This tuple will arrive asynchronously with respect to the operating frequency of the search algorithm.
    % }
    \caption{The communication loop between our asynchronous optimization framework in Lava and the satellite scheduling problem. \textbf{$X'$} is an unknown set of parameters from the search algorithm where the Satellite Scheduling Runtime evaluates \textbf{$X'$} and returns the number of scheduled satellites \textbf{$Y$}. The resulting tuple \textbf{$(X, Y)$} will arrive asynchronously with respect to the configuration, compilation, scheduling, and execution of the satellite scheduler on Loihi.
    }
    \label{fig:system-architecture}
\end{figure}

As shown in Figure~\ref{fig:system-architecture}, our framework introduces an intermediate step between the optimizer and the black-fox function. This step serves multiple functions. (1) It checks for \textit{Stop} or \textit{Pause} commands from the Lava runtime. This serves as a fail-safe to keep the optimizer from running indefinitely. (2) When the optimization process is complete, it will change a boolean flag within the process that can be read through a reference port accessible by the parent process. This handshake operation allows the main thread to know when the asynchronous search process is complete. (3) When a given input port does not have any information, it will put the process to sleep for a specified period of time before probing the port again. If data is available on the port, it will be transmitted to the search algorithm and proceed as expected.

\section{Results \& Discussion}

We evaluate the performance implications of our asynchronous optimization framework by connecting Lava Bayesian Optimization (Lava BO)~\cite{10.1145/3589737.3605998} with a Quadratic Unconstrained Binary Optimization (QUBO) solver applied to a satellite scheduling problem. With Lava BO running on the CPU along with the QUBO solver running on Loihi 2, each time a new set of unknown hyperparameters is generation, it must be transferred to a runtime process. Known as the Satellite Scheduling Runtime, this process has multiple functions to fully evaluate each given set of parameters. (1) The scheduling problem must be initialized with the specification of multiple parameters: the number of satellites, the number of requests, the view height and maximum turning speed of each satellite, the random seed, and the QUBO weights. (2) The solver must be compiled and submitted to the scheduler for allocation and execution on one of the Loihi 2 boards. This process is where synchrony introduces issues within the process graph where the number of time steps required to compile, schedule, and execute the model is a stochastic process. (3) With the results calculated, the parameters and the resulting score are returned to Lava BO where the search process will continue as expected. A visualization of this experiment is shown in Figure~\ref{fig:system-architecture} along with the asynchronous elements. This test scenario highlights the capability of our asynchronous framework to support communicating between multiple processes on different computing architectures where synchrony and runtime determinism are not guaranteed.

\section{Conclusion}

In this work, we present a novel asynchronous optimization and search system within the Lava software framework. With this technology, users can safely interact with processes being executed on different hardware architectures with different operating frequencies. Our system checks whether the input-port has received any spikes. In the case where spikes are not received, the process will go to sleep for a given period of time before probing the port again. This process avoids excess computation where deadlocks will not occur and CPU clock cycles will not be wasted probing ports at full operating frequency.

In future works, we would like to expand the breath of our optimization framework by incorporating multiple agents communicating with a single optimizer. Moreover, we would also like to employ this framework to support lifelong, on-chip learning for robotics and signal processing applications.

% \textcolor{red}{blah blah}

\begin{acks}
The work in this paper is supported by a gift from Intel Corporation.
\end{acks}

\bibliographystyle{ACM-Reference-Format}
\bibliography{sample-base}

\end{document}